%% file: main.tex
\input{preamble.tex}

\begin{document}

\title{Mikado strategy for the detection of atoms in images of microtrap arrays}

\date{\today}

\author{Marc Cheneau}
\email{marc.cheneau@institutoptique.fr}
\author{François Goudail}
\affiliation{Université Paris-Saclay, Institut d'Optique Graduate School, CNRS, Laboratoire Charles Fabry, 91127, Palaiseau, France}

\begin{abstract}
Building on top of our recent work~\cite{Cheneau2025}, we introduce a new strategy to solve the problem of detecting atoms in high-resolution images of microtrap arrays.
By alternating estimation and detection steps, we get rid of the need for an explicit model to compute the posterior occupancy probability of each site given its a priori optimal estimate.
As direct benefits, we show an improved detection accuracy compared to our previous work when the sites are not optically well resolved, and we expect a greater robustness against real experimental conditions.
\end{abstract}

\maketitle

\section{Introduction}

In a recent article~\cite{Cheneau2025}, we introduced a new method to detect the atoms in images of microtrap arrays, which drastically improves the detection accuracy compared to the conventional deconvolution-based approach when neighboring traps are not well resolved optically.
The deconvolution-based approach consists in applying a deconvolution scheme to the measured image, and then in binning the resulting pixel values around the site coordinates to produce an estimate of the "brightness" of each site in the array.
In contrast, our method employs a generalized Wiener filter, which directly inverts the linear relationship between the image and the variables.
It is also optimal in the sense that it represents the brightness of each site and the noise on each pixel by random vectors (represented by \( \mathbold{x} \) and \( \mathbold{n} \), respectively),  and minimizes the mean squared error between the estimated and the true values of the brightnesses.

The mean squared error being a quadratic form, the generalized Wiener filter only depends on the covariance matrices of \( \mathbold{x} \) and \( \mathbold{n} \), which themselves depend on the occupancy probabilities of the sites.
To compute the optimal estimate, one must therefore first determine these probabilities.
In~\cite{Cheneau2025}, we proposed to proceed in two steps.
At the first step, one assumes that all sites have the same occupancy probability.
To determine the best value for this probability, one then maximizes the separation between empty and filled sites in the histogram of estimated brightnesses. This gives the \emph{a priori} optimal estimate.
At the second step, one updates the prior with the information gained at the first step: a low occupancy probability is assigned to the dark sites, while a high occupancy probability is assigned to the bright sites.
One then computes a new estimate using these posterior probabilities, which we call the \emph{a posteriori} optimal estimate.

The a posteriori optimal estimate is always more accurate than the a priori optimal estimate because it uses the information contained in the image to constrain the value of the variables that are estimated.
The difficulty of the estimation problem indeed lies in the coupling between the variables, which itself results from the spatial overlap of the point spread functions (PSF) centered on neighboring sites.
When assigning an occupancy probability close to zero or one to a given site, one effectively constrains its brightness to the vicinity of the expected value for empty or filled sites, respectively.
This constraint effectively decouples the variable from the surrounding variables, which can then be estimated more accurately.

While the a posteriori optimal estimator can drastically improve the detection accuracy compared to the deconvolution-based approach, it still has an intrinsic limitation; it is the need for an explicit model to deduce the posterior occupancy probabilities from the a priori optimal estimate.
Such model can be difficult to find, especially when the overlap between the PSF of neighboring sites induces anti-correlations between the estimation errors.
In the present article, we present an alternative strategy which does not require such model.
As a consequence, it further improves the detection accuracy in the regime of strong spatial overlap between the PSFs of neighboring sites, and is also expected to be more robust in a real experimental context.
The basic idea is to alternate estimation and detection steps, such that the problem gets progressively simplified.
We call this strategy the "mikado" strategy, in reference to the classic pick-up sticks game.
We also provide the reader with a tutorial example in the form of a JUPYTER notebook~\cite{Notebook}.

In the following we remind the reader of the formalism and model behind the generalized Wiener filter (\cref{sec:wiener_filter}), explain the working principle mikado strategy (\cref{sec:mikado}), and compare its performance to both the two-step strategy and the method based on the Wiener deconvolution (\cref{sec:benchmarking}). Finally, we illustrate the potential of our new method by considering a recent experiment with Erbium atoms in an optical lattice \cite{Su2023,Su2025} (\cref{sec:use_case}).

\section{Generalized Wiener filter}%
\label{sec:wiener_filter}

The starting point of the generalized Wiener filter is the linear relationship between the measured signal and the variables, which can be written in matrix form as
\begin{equation}
    \label{eq:model}
    \mathbold{y} = \mathbold{Mx} + \mathbold{k} + \mathbold{n} \; .
\end{equation}
Here, \( \mathbold{y} \) is a vectorized representation of the image, \( \mathbold{x} \) is the vector of site brightnesses, \( \mathbold{k} \) is the vector of background, and \( \mathbold{n} \) the vector of noise.
Each column of the measurement matrix \( \mathbold{M} \) represents the image of one site in array, normalized to unity.
In the following we will implicitly subtract the background from the image, and replace \( \mathbold{y} - \mathbold{k} \) with \( \mathbold{y} \).

The generalized Wiener filter is the matrix \( \mathbold{H} \) which minimizes the mean squared error between the estimated and the true values of the variables:
\begin{equation}
    \label{eq:argmin}
    \mathbold{H}_\text{opt} \triangleq \argmin_{\mathbold{H}} \brk[s]{ \text{MSE}\brk{\mathbold{H}} } \; ,
\end{equation}
with
\begin{align}
    \label{eq:MSE}
    \text{MSE}(\mathbold{H}) & \triangleq \avg1{ \norm{ \hat{\mathbold{x}}(\mathbold{H}) - \mathbold{x}} }^2 \; , \\
    \label{eq:inverse_problem}
    \hat{\mathbold{x}}(\mathbold{H}) & \triangleq \mathbold{H} \brk{\mathbold{y} - \mathbold{M}\avg{\mathbold{x}}} + \avg{\mathbold{x}} \; .
\end{align}
Here, \( \norm{\cdot} \) is the Euclidean norm, and \( \avg{\cdot} \) denotes the expectation value over the probability distributions of \( \mathbold{x} \) and \( \mathbold{n} \), which are treated as random vectors.
The expected value of \( \mathbold{x} \) is subtracted to ensure that the estimator is unbiased.

The solution to \cref{eq:argmin} is
\begin{equation}
    \label{eq:OLE}
    \mathbold{H}_\text{opt} = \brk{\mathbold{M}^\intercal \mathbold{\Sigma}_n^{-1} \mathbold{M} + \mathbold{\Sigma}_x^{-1}}^{-1} \mathbold{M}^\intercal \mathbold{\Sigma}_n^{-1} \; ,
\end{equation}
where \( \mathbold{\Sigma}_x \) is the covariance matrix of the variables, and \( \mathbold{\Sigma}_n \) is the covariance matrix of the noise.
In practice, rather than computing \( \mathbold{H}_\text{opt} \) explicitly, one solves the linear system
\begin{subequations}%
    \label{eq:linear_system_full}
    \begin{equation}
        \label{eq:linear_system}
        \mathbold{A} \brk{\hat{\mathbold{x}} - \avg{\mathbold{x}}} = \mathbold{B} \brk{\mathbold{y} - \mathbold{M} \avg{\mathbold{x}} }
    \end{equation}
    for \( \mathbold{x} \), with
    \begin{align}
        \mathbold{A} &= \mathbold{M}^\intercal \mathbold{\Sigma}_n^{-1} \mathbold{M} + \mathbold{\Sigma}_x^{-1} \; , \\
        \mathbold{B} &= \mathbold{M}^\intercal \mathbold{\Sigma}_n^{-1} \; .
    \end{align}
\end{subequations}
The preferred method to compute the solution of \cref{eq:linear_system_full} is the conjugate gradient method with the Crout version of the incomplete {LU} decomposition as a preconditioner.
The dropping rules applied to ensure the sparsity the L and U matrices should be adapted to the amount of overlap between the PSF of neighboring sites: a large overlap requires more non-zero entries than a small overlap.

We assume that each site has an occupancy probability \( \brk{\mathbold{p}}_i \), and that the brightness of filled sites fluctuates around an expected value \( \mu \) with a variance \( \sigma^2 \).
These fluctuations account for site-to-site differences in the trap depth or in the illumination, for instance.
The mean and covariance matrix of \( \mathbold{x} \) are then given by
\begin{align}
    \label{eq:meanx}
    \brk{\avg{\mathbold{x}}}_i
    &= \brk{\mathbold{p}}_i \mu \; , \\
    \label{eq:covx}
    \brk{\mathbold{\Sigma}_x}_{ij}
    &= \brk[c]1{ \brk{\mathbold{p}}_i \brk[s]{1 - \brk{\mathbold{p}}_i} \mu^2 + \brk{\mathbold{p}}_i \sigma^2 } \delta_{ij} \; ,
\end{align}
If the variance for a given site is small, i.e. \( \brk{\mathbold{\Sigma}_x}_{ii} \ll \mu^2\), the estimate will be constrained to the vicinity of the expected value \( \avg{\brk{\mathbold{x}}_i} \).
This is the case when \( \brk{\mathbold{p}}_i \simeq 0 \) or 1.
On the contrary, if the variance is large, i.e. \( \brk{\mathbold{\Sigma}_x}_{ii} \sim \mu^2\), the estimate will be free to adapt to the image.
This is the case when \( \brk{\mathbold{p}}_i \simeq 0.5 \).

The noise on each pixel consists of three contributions: the shot noise of the atomic signal, with a variance \( \mu \brk{\mathbold{M p}}_i \), the shot noise of the background, with a variance \( \brk{\mathbold{k}}_i \), and the camera read noise, with a variance \( r^2 \).
The covariance matrix of the noise is therefore
\begin{equation}
    \label{eq:covn}
    \brk{\mathbold{\Sigma}_n}_{ij} = \brk[s]1{  \mu \brk{\mathbold{M p}}_i + \brk{\mathbold{k}}_i + r^2} \delta_{ij} \; .
\end{equation}
The mean of the noise is zero on every pixel.

\section{The mikado strategy}%
\label{sec:mikado}

We designed the mikado strategy to circumvent a limitation inherent to the two-step method introduced in \cite{Cheneau2025}, namely the need for an explicit model to compute the posterior occupancy probability of each site given its a priori optimal estimate.
The accuracy of such model is especially important when the PSF of neighboring sites strongly overlap, and the two modes in the histogram are no longer well separated.
A typical choice is to use a mixture model, such as a Gaussian mixture model, which treats the estimated brightnesses for each site in the array as independent random variables.
The issue with such model is that it overlooks the anti-correlations between the estimation errors at neighboring sites, which stem from the spatial overlap between the PSFs.

\begin{figure}
    \centering
    \includegraphics{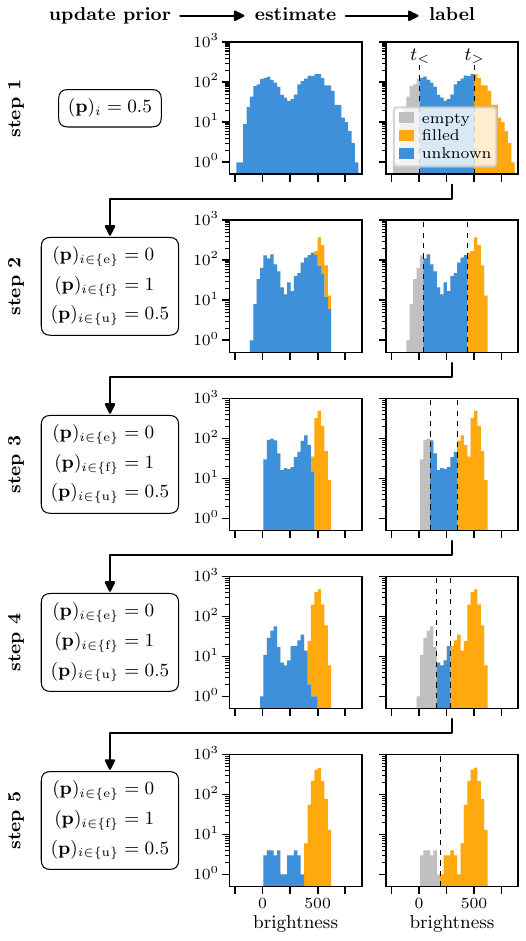}
    \caption{Working principle of the mikado strategy.
    The method is applied to a simulated image corresponding to a square array of \numproduct{50 x 50} sites with a uniform occupancy probability \( p = \num{0.6} \).
    Each step consists of three phases: the prior update, the estimation, and the classification.
    Using two thresholds \( t_< \) and \( t_> \) indicated by the dashed vertical lines, the sites are classified as empty (abbreviated "e", gray bins), filled ("f", yellow bins), or with unknown occupancy ("u", blue bins).
    Empty sites are not included in the estimation since they do not contribute to the signal.
    The parameters used to generate the image are as described in \cref{sec:benchmarking}, with \( \mu = 500 \) and \( a = \num{1.25}\,r_\text{PSF} \).
    With these parameters, the detection error rate of the mikado strategy is \qty{0.1 +- 0.1}{\percent}, where the uncertainty is the standard deviation over \num{100} similar images.
    As a comparison, the detection error rate of the method based on the Wiener deconvolution is \qty{2.6 +- 0.4}{\percent}, that of the a priori optimal estimator is \qty{2.0 +- 0.3}{\percent}, and that of the a posteriori optimal estimator is \qty{0.5 +- 0.2}{\percent}.
    }%
    \label{fig:mikado}
\end{figure}

The mikado strategy proceeds by successive simplification of the estimation-detection problem, as illustrated in \cref{fig:mikado}.
At each step, one updated the prior, computes a new estimate, and label the sites as empty, filled or unknown, depending on how their brightness compare with two thresholds \( t_< \) and \( t_> \), with \( t_< \leq t_> \).
The prior update proceeds according to the labels given at the previous step:
The empty sites are simply eliminated from the problem by removing the corresponding rows and columns in the linear system \labelcref{eq:linear_system_full}.
The filled sites remain among the variables, but with an occupancy probability of 1, and hence little freedom to depart from the expected value \( \mu \).
Finally, the sites whose occupancy is still unknown get an occupancy probability of \num{0.5}, which gives them the most freedom to adapt to the measured signal.
For the first step, we simply assume a uniform occupancy probability of \num{0.5}.

The convergence of the sequence is enforced by moving \( t_< \) and \( t_> \) towards each other until they coincide.
The exact progression of these thresholds does not matter as long as the probability to incorrectly classify a site as empty or filled is small at each step.
This relative insensitivity to the parameterization is illustrated by \cref{fig:tuning_params} in \cref{app:tuning_params} on a concrete example.
In this work, we always set \( t_< = 0 \) and \( t_> = \mu \) for the first step, \( t_< = t_> = 0.4\ \mu \) for the last step, and set the number of steps to either 5 or 10, depending on which of the computation time or the accuracy has the most importance.

\section{Benchmarking}%
\label{sec:benchmarking}

We benchmark the mikado strategy against both the a priori and a posteriori optimal estimators, and against the estimator based on the Wiener deconvolution (the latter is described in \cite{Cheneau2025}).
Each estimator is applied to simulated test images of a square array of \numproduct{50 x 50} sites with a uniform occupancy probability \( p = \num{0.6} \).
The brightness of filled sites follows a normal distribution with mean \( \mu \) and standard deviation \( \sigma = \mu / 10 \). 
The point spread function is a Gaussian, with half width at half maximum \( r_\text{PSF} = \qty{2}{pixels} \).
The distance between neighboring sites is denoted \( a \).
We assume a zero background for simplicity, and take the camera read noise to be normally distributed around zero with a standard deviation \( r = 1 \).

\begin{figure}
    \centering
    \includegraphics{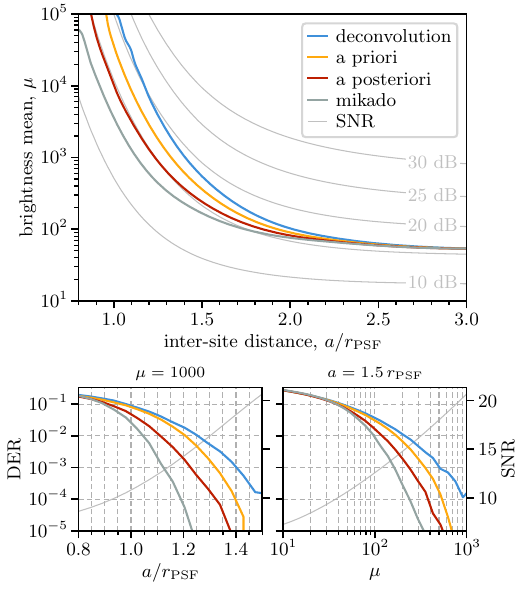}
    \caption{Detection error rate (DER) and signal-to-noise ratio (SNR) as a function of the brightness mean and ratio of the inter-site distance to the PSF half width at half maximum.
    Top panel: The thick colored lines are the loci of the points where the DER of each estimator is equal to \( 10^{-3} \).
    The thin gray lines are the SNR level curves at \qtylist{10;15;20;25;30}{\dB}, see~\cite{Cheneau2025}.
    Lower panels: Variations of the DER with \( a \) for \( \mu = 10^3 \) (left), and with \( \mu \) for \( a = 1.5 \, r_\text{PSF} \) (right). The thick colored lines represent the DER, and the thin gray lines represent the SNR\@.
    The DER is averaged over 100 images.
    The LU decomposition is parameterized with a threshold of \( 10^{-3} \) and a fill-in of \( 10^3 \).
    The mikado estimator proceeds in 10 steps.
    }%
    \label{fig:der}
\end{figure}

We start by comparing the detection error rate (DER) of each estimator.
The DER is defined as the number of false positives and false negatives divided by the number of sites; its value therefore depends on the level of the detection threshold.
Here, we set the detection threshold at the level which minimizes the DER~\footnote{In the case of the mikado estimator, the sequence of thresholds still goes from \( t_\lessgtr^{(1)} = (0,\,\mu) \), to \( t_\lessgtr^{(N)} \triangleq t^{(N)} = 0.4\,\mu \), but the labelling at the end of the last step is done using the optimized detection threshold, rather than \( t^{(N)} \), which is only possible of course because we know the ground truth.
This is not in the spirit of the mikado estimator of course, but we do so to enable a fair comparison with the other estimators.}.
\cref{fig:der} illustrates how the DER of each estimator depends on \( a \) and \( \mu \).
The top panel shows the locus of the points where the DER is equal to \( 10^{-3} \).
The thin gray lines indicate level curves for the signal-to-noise ratio (SNR) introduced in~\cite{Cheneau2025}.
The lower panels in \cref{fig:der} show two cuts through the plane \( \pair{a}{\mu} \): one along the line \( \mu = 10^3 \) (left) and one along the line \( a = 1.5 \, r_\text{PSF} \).
We distinguish two regimes:
When the sites are well resolved (\( a \gg r_\text{PSF} \)), all estimators display the same detection accuracy, which is solely determined by the noise in the image.
When the sites are not well resolved (\( a \lesssim r_\text{PSF} \)), the mikado estimator outperforms all other estimators because of its greater ability to untangle the variables for a given noise level.

\begin{figure}
    \centering
    \includegraphics{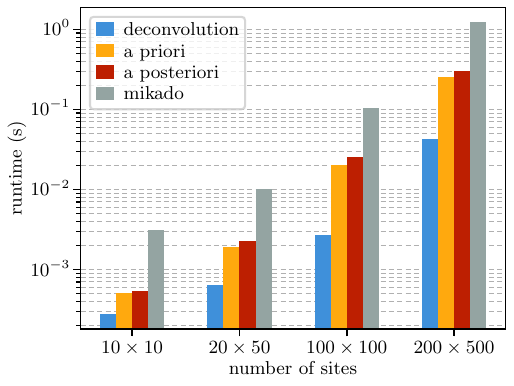}
    \caption{Runtime comparison.
    The runtime is the averaged over 10 test images and 10 repetitions for each image.
    We adapt the number of nonzero entries to keep in the LU matrices to the ratio \( a/r_\text{PSF} \) by checking the convergence of the CG algorithm.
    The conjugate gradient algorithm stops when the relative difference between the left and right-hand side of \cref{eq:linear_system} is less than \( 10^{-2} \).
    The LU decomposition is parameterized with a threshold of \( 10^{-2} \) and a fill-in of \( 10^2 \).
    The mikado strategy proceeds in 5 steps.
    }%
    \label{fig:runtime}
\end{figure}

Next, we measure the computation times of the different estimators for an increasing number of sites in the array.
The computation time of the estimators based on the generalized Wiener filter (a priori optimal, a posteriori optimal and mikado) is dominated by the computation of the incomplete LU factorization, and mostly depends on the ratio \( a / r_\text{PSF} \) (one must increase the number of non-zero entries in the L and U matrices when this ratio decreases).
In contrast, the computation time of the method based on the Wiener mostly depends on the size of the image.
The results obtained for \( a = \num{1.25}\, r_\text{PSF} \) and \( \mu = 500 \) are summarized in \cref{fig:runtime} 
Here, we have parameterized the mikado estimator to proceed in 5 steps, resulting in a computation time which is roughly 5 times longer than that of the a priori optimal linear estimate.
We still regard the mikado estimator as being fast, since current experiments operate on arrays with less than \( 10^4 \) sites, for which the runtime of the mikado estimator is at most of the order of \qty{100}{\ms}.

\section{Practical use case}%
\label{sec:use_case}

To illustrate the potential of our method, we consider the recent demonstration of site-resolved imaging of an Erbium gas in a tunable-spacing accordion lattice~\cite{Su2023,Su2025}.
In this experiment, the inter-site spacing is \qty{266}{\nm}, the imaging wavelength is \qty{401}{\nm}, and the numerical aperture of the microscope objective is \num{0.85}.
The strategy reported by the authors to solve the detection problem was experimental, rather than algorithmic: just before imaging, they dilated the lattice spacing to a few micrometers by reducing the angle at which the lattice beams were crossing.
With such a large spacing, it was possible to detect the atoms with a high accuracy using a very short exposure time, without having to care about trapping and laser cooling during imaging.
Still, we can look whether it would be possible in principle to detect the atoms directly in the lattice with a \qty{266}{\nm} spacing.
With \( a = \qty{266}{\nm} \) and \( r_\text{PSF} \) given by the half width at half maximum of a diffraction-limited PSF, we get \( a / r_\text{PSF} \simeq \num{1.1} \).
We can read from \cref{fig:der} that it would be necessary to collect about \num{1000} photoelectrons per atom on the camera to achieve a DER of \( 10^{-2} \).
For more standard atomic species, collecting such a number of photoelectrons is not a big experimental issue. Provided that this is also the case for Erbium, we conclude that the implementation of a tunable-spacing accordion lattice may not be a strict necessity for imaging.

\section{Conclusion}

We have described a model-free, iterative strategy to solve the problem of detecting atoms in high-resolution images of microtrap arrays.
By alternating estimation and detection steps, it progressively reduces the correlations between estimation errors, which results in an improved detection accuracy compared to the existing methods.
Being model-free, we expect that it will be more robust against real experimental conditions.
This new method represents a further step towards a more rational and efficient solution
to the atom detection problem, with direct benefits to existing and future experiments.

\begin{acknowledgments}
    The authors are grateful to Romaric Journet for the critical reading of this manuscript.
\end{acknowledgments}

\appendix
\crefalias{section}{appendix}

\section{Optimization of the mikado sequence}%
\label{app:tuning_params}

\begin{figure}
    \centering
    \includegraphics{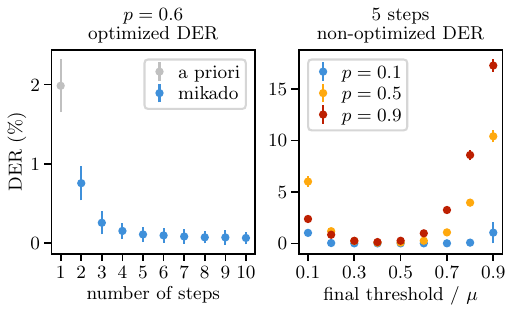}
    \caption{Influence of the number of steps (left) and of the final threshold position (right) on the detection error rate (DER) of the mikado strategy.
    Both measures are performed with \( \mu = 500 \), \( \sigma = \mu / 10 \), \( r_\text{PSF} = \qty{2}{pixels} \), \( a = \num{1.25}\,r_\text{PSF} \), \( k = 0 \) and \( r = 1 \).
    The error bars represent the standard deviation over a set of \num{100} images.
    }%
    \label{fig:tuning_params}
\end{figure}

The mikado sequence is parameterized by the sequence of thresholds \( \set{\pair{t_<^{(n)}}{t_>^{(n)}}}_{n = 1}^{N} \), with
\begin{equation}
    t_\lessgtr^{(n)} = t_\lessgtr^{(1)} + \frac{n - 1}{N - 1} (t^{(N)} - t_\lessgtr^{(1)}) \;,
\end{equation}
and \( t_<^{(1)} < t^{(N)} < t_>^{(1)} \).
\cref{fig:tuning_params} illustrates the influence of the number of steps \( N \) (left) and of the final threshold position \( t^{(N)} \) (right) on the detection error rate (DER).
The DER is measured by applying a detection threshold to the estimate produced at the last step.
The spirit of the mikado estimator is to use \( t^{(N)} \), and this is what we do for the graph on the right.
For the graph on the left, however, we rather choose the detection threshold which minimizes the DER to enable a fair comparison with the a priori optimal estimate.

\bibliography{bibliography.bib}

\end{document}

%% file: preamble.tex
\documentclass[aps,prapplied,twocolumn,floatfix]{revtex4-2}

\usepackage{mathtools,amsfonts,amssymb}
\usepackage{fixmath}
\usepackage{bm}
\usepackage{nccmath}
\usepackage{xfrac}
\usepackage{xspace}
\usepackage[extdef]{delimset}
\usepackage{graphicx}
\usepackage{xcolor}
\usepackage[inline,shortlabels]{enumitem}
\usepackage{calc}
\usepackage{soul}
\usepackage{hyperref}
\usepackage[capitalize]{cleveref}
\usepackage{siunitx}
\usepackage[casesstyle]{cases}
\usepackage{booktabs,multirow}
\usepackage{csquotes}
\MakeOuterQuote{"} 

\DeclareMathOperator*{\argmin}{arg\,min}